\documentstyle[prl,aps,epsbox]{revtex}
\begin{document}
\draft
\preprint{}
\twocolumn[\hsize\textwidth\columnwidth\hsize\csname@twocolumnfalse%
\endcsname 
\title{ Double Spiral Energy Surface \\
in One-dimensional Quantum Mechanics
of Generalized Pointlike Potentials }
\author{
Taksu Cheon}
\address{
Laboratory of Physics, Kochi University of Technology,
Tosa Yamada, Kochi 782-8502, Japan \\
email: cheon@mech.kochi-tech.ac.jp, 
    http://www.mech.kochi-tech.ac.jp/cheon/
}
\date{May 8, 1998}
\maketitle
\begin{abstract}
We analyze the eigenvalue problem of a quantum particle on the line 
with the generalized pointlike potential of three parameter family.
It is shown that the energy surface in the parameter space has a set
of singularities, around which different eigenstates are connected
in the form of paired spiral stairway.
An exemplar wave-function aholonomy is displayed where the ground 
state is adiabatically turned into the second excited state 
after cyclic rotation in the parameter space. 

\vspace*{3mm}
\end{abstract}
%
]
\narrowtext

%
%
The evolution of physical states under the cyclical variation of 
environmental variables has been one of the key concepts of 
thermodynamics and its engineering applications.
That concept has entered into the quantum mechanics with the
discovery of Berry phase \cite{BE84,AA87}
and its subsequent non-Abelian generalization \cite{WZ84}. 
A quantum eigenstate of parametric Hamiltonian
can be adiabatically turned into a different state belonging to 
the same energy multiplet after the cyclic change of the parameters.
The discovery has had profound impacts in the various fields of 
theoretical physics.  Most notable among them is its
implication on the origin of the gauge field.

Underlying the phenomena of Berry phase is the structure of 
the energy surface in the parameter space in the form of two cones, 
one upside-down which touch each other at the apices.
If there can be other non-trivial structure than the double cones, 
that could lead to the wave-function aholonomy
and gauge field structure which are 
more exotic than the Berry phase phenomena.

In this paper, we report the finding of just such 
non-trivial energy surface in parameter space in one of the simplest
conceivable quantum system, that of a one-dimensional particle
subject to a potential which is zero except at a single point.
The energy surface of the system is characterized by a singularity
which acts as a branch point that connects different energy levels 
in the form of double spiral.
By the cyclical variation of the parameters 
around the singularity, one can adiabatically transform, for example, 
the ground state into the second excited state.

%
%
%
Let us start by introducing our model system.
The most general pointlike potential in one dimension 
with time-reversal symmetry is of three parameter family
\cite{GK85,SE86,SE86a,AG88}. It has been something of a
mystery until
its explicit construction in terms of self-adjoint local
operator has been devised \cite{CS98}.  Here, we
outline the procedure. 
We define a function consisting of three 
Dirac's deltas placed next to each other in one dimension;
\begin{eqnarray} 
\label{01} 
\xi(x; v_-, u, v_+ ; a) 
= v_- \delta(x+a) + u \delta(x) + v_+\delta(x-a).
\end{eqnarray}
We let the strengths $v_-$, $v_+$ and $u$ to vary with the distance
$a$ in the form
\begin{eqnarray} 
\label{02} 
v_-(a) &=& -{1 \over {2a}}+ {{\gamma -1} \over {2 \delta}},
\\ \nonumber 
v_+(a) &=& -{1 \over {2a}}+ {{\alpha -1} \over {2 \delta}} 
\\ \nonumber 
u(a) &=& -{1 \over {a}} - {{\alpha\gamma -1} \over {2\beta a^2}} 
\end{eqnarray}
where $\alpha$, $\beta$, $\gamma$ and $\delta$ are the real numbers
which satisfy the constraint
\begin{eqnarray}
\label{3}
\alpha \gamma -\beta \delta =1.
\end{eqnarray}
The zero distance limit of the function $\xi (x)$ defines the generalized
pointlike potential
\begin{eqnarray} 
\label{03} 
\chi(x; \alpha, \beta, \gamma, \delta)
\equiv \mathop {\lim}\limits_{a\to 0}\xi (x;v_-(a),u_0(a),v_+(a);a) ,
\end{eqnarray}
which, when used in the Schr{\" o}dinger equation
\begin{eqnarray} 
\label{1} 
\left[
-{1 \over 2}{{d^2} \over {dx^2}}
+\chi (x;\alpha ,\beta ,\gamma ,\delta )
\right] \psi (x)
=E(\alpha ,\beta ,\gamma ,\delta )\psi (x) ,
\end{eqnarray}
yields the eigenfunction $\psi (x)$ that
has discontinuity at $x = 0$ both in $\psi (x)$ itself
and in its derivative $\psi '(x)$, whose amount is specified by
\begin{eqnarray}
\label{2}
\psi'(0_+)+\alpha \psi'(0_-)=-\beta \psi(0_-)
\\ \nonumber
\psi(0_+)+\gamma \psi(0_-)=-\delta \psi'(0_-).
\end{eqnarray}
The constraint, Eq. (\ref{3}) guarantees 
the self-adjointness of the Hamiltonian 
operator of Eq. (\ref{1}),
which is equivalent, in physical term,
to the requirement of the continuity of the
probability flux at $x = 0$.

In order to make the quantum spectra discrete, we place
the system in a finite region on the line
(``one-dimensional billiard'') of the length $L$, 
depicted in Fig. I, by imposing
the wave functions to disappear at $x=-rL$ and $x=(1-r)L$.

\begin{figure}
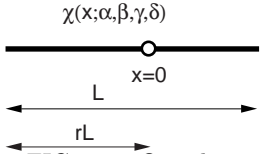

\label{fig1}
\center\psbox[hscale=0.45,vscale=0.45]{spfig1.epsf}
\caption{
One dimensional line of length $L$ with a point potential at $x=0$
which divides the line into two parts with lengths $rL$ and $(1-r)L$.
} 
\end{figure}
Thanks to the zero-range nature of the interaction,
the system is solvable in the sense that the resolvent 
is explicitly calculable.
We opt for an elementary derivation.  The positive energy
solution of Eq. (\ref{1}) is written in the form 
\begin{eqnarray}
\label{4}
\psi (x)&=&A_+\sin k(x-L+rL) \ \ \ (x>0) ,
\\ \nonumber
\psi (x)&=&A_-\sin k(x+rL) \ \ \ (x<0) ,
\end{eqnarray}
where $k = \sqrt{2E}$ is the linear momentum.  
For the negative energy case $E < 0$, one makes the replacement
\begin{eqnarray}
\label{5}
\sin \to \sinh, \ \ 
\cos \to \cosh, \ \
k \to \kappa. 
\end{eqnarray}
where $\kappa = \sqrt{-2E}$.
From Eqs.(\ref{2}) and (\ref{4}), one obtains the eigenvalue
equation for the case of $E > 0$ in the form
\begin{eqnarray}
\label{6}
F(k) = 0
\end{eqnarray}
where 
\begin{eqnarray}
\label{7}
F(k)&=&\alpha \sin k(1-r)L\cos krL
\\ \nonumber
& &+\gamma \cos k(1-r)L\sin krL
\\ \nonumber
& &+\beta {1 \over k}\sin k(1-r)L\sin krL
\\ \nonumber
& &+\delta k\cos k(1-r)L\cos krL .
\end{eqnarray}
The corresponding eigenfunction is obtained from the relation
\begin{eqnarray}
\label{8}
{{A_+} \over {A_-}}
&=&\alpha \cos k(1-r)L\cos krL
\\ \nonumber
& &-\gamma \sin k(1-r)L\sin krL
\\ \nonumber
& &+\beta {1 \over k}\cos k(1-r)L\sin krL
\\ \nonumber
& &-\delta k\sin k(1-r)L\cos krL.
\end{eqnarray}
The negative eigenvalue (when it exists) can be obtained in a
parallel manner with the replacement Eq. (\ref{5}).

Since visualizing the four-dimensional energy surface 
is out of our usual intuition, we look at the eigenvalues as
functions of two parameters $( \alpha, \beta)$ with
a fixed value  $\gamma = \gamma_0$, namely,
\begin{eqnarray}
\label{9}
E(\alpha ,\beta )
\equiv E(\alpha ,\beta ,\gamma =\gamma _0,
\delta = \frac{\alpha \gamma _0-1}{\beta }) .
\end{eqnarray}
A curious feature of the energy surface Eq. (\ref{9}) is that
the value of $\delta$ becomes indefinite for a special point
in the parameter space
$(\alpha^* , \beta^*) = (1/\gamma_0 , 0)$.
Since the function $F(k)$, Eq. (\ref{7})
is a smooth function of
all parameters $(\alpha , \beta , \gamma , \delta )$,
this implies that the energy
eigenvalue is also indefinite at this point, namely
\begin{eqnarray}
\label{10}
E(\alpha^*,\beta^*) \ \ {\rm indefinite \ for} \ \  
(\alpha^*,\beta^*)=({1 \over {\gamma _0}},0) .
\end{eqnarray}
We look at the neighbourhood of the singularity by
defining the polar coordinate  $(\rho ,\theta )$ 
around the singularity;
\begin{eqnarray}
\label{11}
\alpha &=&{1 \over {\gamma _0}}-\rho \sin (\theta )
\\ \nonumber
\beta &=&\rho \cos (\theta ).
\end{eqnarray}
The equation Eq. (\ref{7}) takes the form
\begin{eqnarray}
\label{12}
F(k) &=&
{1 \over {\gamma _0}}\sin k(1-r)L\cos krL
\\ \nonumber
& &+\gamma _0\cos k(1-r)L\sin krL
\\ \nonumber
& &-\gamma _0 k\cos k(1-r)L\cos krL\tan \theta + O(\rho) .
\end{eqnarray}
Neglecting the term of $O(\rho)$, we can express the 
solution of Eq.(\ref{6}) analytically as
\begin{eqnarray}
\label{13}
\theta (k) =
\arctan \left( {{1 \over {k{\gamma _0}^2}}
\tan k(1-r)L+{1 \over k}\tan krL} \right) .
\end{eqnarray}
%
\begin{figure}
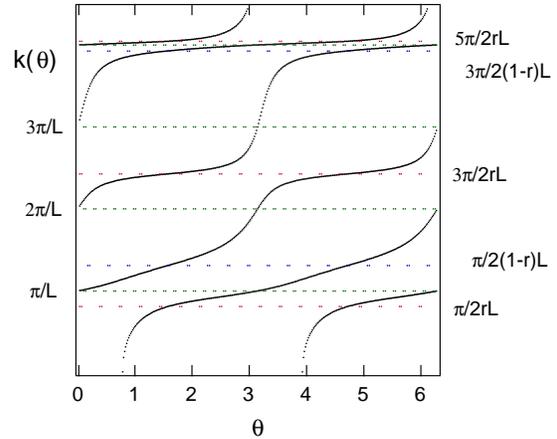

\label{fig2}
\center\psbox[hscale=0.45,vscale=0.45]{spfig2.epsf}
\caption{
The solution Eq. (13) of the Scr{\" o}dinger equation 
around the singularity at $(\alpha , \beta) = (1/\gamma_0 , 0)$.
The $\gamma_0$ is chosen to be $-1$. Other parameters are 
$L=1$ and $r = 0.618034$. 
} 
\end{figure}
This function is multivalued.  
But at each branch $\theta_n(k)$, it satisfies
$d\theta_n / dk > 0$, therefore can be inverted 
to obtain an increasing function $k_n(\theta)$.
On the other hand, 
$k(\theta)$ as a whole is periodic with the period $\pi$,
reflecting the periodicity of $F(k)$.
These two facts can be made compatible only when each branch
$k_n(\theta)$ is connected to the next branch 
after moving up for the period $\pi$.
Same hold for the function $\kappa (\theta)$ bellow zero energy.
A notable fact is that one has $\kappa \to \infty$ when
$\theta \to 0_+$, which means that a bound state emerge from
negative infinite energy at $\theta = 0_+ , \pi _+ , 2 \pi _+ , ...$.  
The situation can be best understood by 
inspecting the Fig. II where the function 
$k(\theta)$ is drawn for
the case of $L = 1$, 
$r = (\sqrt{5}-1)/2 \approx 0.618034$ 
and $\gamma_0 = -1$.
We now easily see the remarkable property of
the energy eigenvalue around the singularity,
which can be expressed as
\begin{eqnarray}
\label{14}
E_n(\theta +\pi )=E_{n+1}(\theta ) ,
\end{eqnarray}
and therefore
\begin{eqnarray}
\label{15}
E_n(\theta +2\pi )=E_{n+2}(\theta ) .
\end{eqnarray}
Note the fact that while Eq. (\ref{14}) is
an approximation based on the expansion Eq. (\ref{12}),
the relation Eq. (\ref{15}) is an exact one. 
%
%
%
\begin{figure}
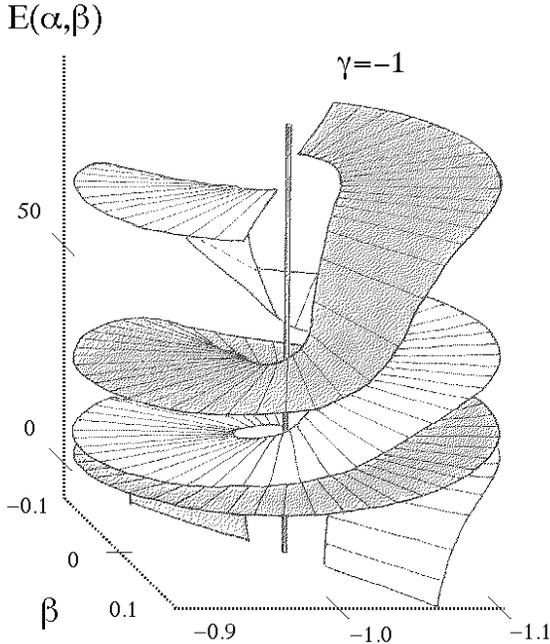

\label{fig3}
\center\psbox[hscale=0.45,vscale=0.45]{spfig3.epsf}
\caption{
Energy surface $E( \alpha, \beta )$ with a fixed value for 
$\gamma =\gamma_0 = -1$.  Other parameters are 
$L=1$ and $r = 0.618034$.
}
\end{figure}
Although essential features of the energy surface
around the singularity is expressed 
in Eq. (\ref{13}) and Fig. II, it is helpful 
to look at the full energy surface $E(\alpha, \beta)$
around the singularity, Eq. (\ref{10}),
which is depicted in Fig. III.
The values for $r$ and $\gamma_0$ are the same as in Fig. II.
But here, the energy is calculated directly from
Eqs. (\ref{6}), (\ref{7}) without any approximation.
The spiral structure is clearly discernible
from this figure.  Because of the approximate periodicity of
$E(\alpha, \beta)$ by the period $\pi$ as expressed in
Eq. (\ref{14}), the structure is approximately axis-symmetric 
with respect to the singular axis.  The resulting structure is
a {\it double spiral stairway},
somewhat reminiscent of the celebrated double helix of DNA
in its appearance.

Next, we take a brief look at the gauge field structure 
behind the scene.
In the manner of Wilczek and Zee\cite{WZ84},
adopting the vector notation $\vec\lambda = (\alpha, \beta)$, 
we consider the evolution of the eigenstate $\phi_n(\vec\lambda)$ 
with the adiabatic variation of the parameter $\vec\lambda$ that
start from $\vec\lambda_0$, at which point 
we set $\phi_n(\vec\lambda_0)$ $= \psi_n(\vec\lambda_0)$.
We then have 
\begin{eqnarray}
\label{16}
\phi_n(\vec\lambda) 
= \sum_{m} {U_{n,m}(\vec\lambda ,\vec\lambda_0) 
           \psi_m(\vec\lambda_0)} 
\end{eqnarray}
with the evolution matrix $U$ given by the path-ordered integral
\begin{eqnarray}
\label{17}
U(\vec\lambda ,\vec\lambda_0) 
= P \exp {\int_{\vec\lambda_0}^{\vec\lambda} 
          \vec{A}(\vec\lambda)\cdot d\vec\lambda}
\end{eqnarray}
where $\vec{A}$ is the Berry-Mead connection \cite{BE84,ME79}
defined by
\begin{eqnarray}
\label{18}
\vec{A}_{n,m}(\vec\lambda) 
= \left< \psi_n (\vec\lambda)\right|
  {\partial\over\partial\vec\lambda}
  \left. \psi_m (\vec\lambda)\right>.
\end{eqnarray}
The non-trivial nature of the connection $\vec{A}(\vec\lambda)$
is evident in the evolution integral over a closed-loop $C$
\begin{eqnarray}
\label{19}
U_{n,m}(C) &=& \delta_{n,m+2} 
\ \ \ \ {\rm if} \  C \  {\rm encircles} \  
         \vec\lambda^* =(\alpha^*, \beta^*) 
\\ \nonumber 
          &=& \delta_{n,m} 
\ \ \ \ \ \ \ \ {\rm otherwise} ,
\end{eqnarray}
which is just another expression of
the spiral structure, Eq. (\ref{15}).

It is instructive to look at the way an eigenfunction 
at a given parameter value $(\alpha ,\beta)$ 
is turned into the second-next higher state after 
the rotation around the singularity.  
In Fig. IV, an example of such wave-function 
morphology is displayed.
One essential feature is the node change which occurs
at the crossing of $\beta = 0$ line.  On this line, the
system is separated into the non-communicating two
subsystems at $x = 0$ because of the infinite value of $\delta$.
The reason of nodal change is that the approach to $\beta = 0$ 
line from the $\beta >0$ side and $\beta < 0$ side
are physically distinct because they each
correspond to $\delta = \pm \infty$ and $\delta = \mp \infty$
respectively.
(The composite signs are for the case of $\alpha \gamma_0 >1$
and $\alpha \gamma_0 <1$ respectively, in turn.)
Yet, at $\beta = 0$ line, two cases get connected smoothly.

It is easy to see that the energy surface has a similar
structure after interchanging the role of 
$\beta$ and $\delta$.  One therefore has
the singularity in $(\alpha , \delta)$ space at the point
$(1/\gamma_0 , 0)$.  Also one can interchange the parameter set 
$(\alpha , \gamma)$ and $(\beta , \gamma)$ to get the singularity
in $(\gamma , \delta)$ space, etc..  It would be very interesting
if one could represent the full four-dimensional energy surface
in intuitive fashion, although doing so would require 
a real ingenuity. 

%
%
We now place our findings in broader context.
Looking at the spiral structure of Fig. IV, one is inevitably 
reminded with the Bender-Wu singularity which is found 
in the energy surface of a {\it complex} parameter of 
the quartic quantum oscillator \cite {BW69}.
Although we do not find any direct link between our problem and
the quartic oscillator, the similarity is still intriguing.
 
It is natural to search for other quantum systems whose
parametric energy surface has similar structure.  
Whether the new type of aholonomy found in our example 
is as ubiquitous as the Berry phase aholonomy is an 
important question.
It is known that the contact interactions on one-dimensional
line have immediate generalization as the contact interactions 
{\it on the graph} \cite{EX96}.  That could be one place where 
one might find more examples of interesting wave-function
morphology.
Also, a recent work on ``homeopathic'' quantum system \cite{AB97}
seems to have certain relevance to the aholonomy discussed here.

\begin{figure}
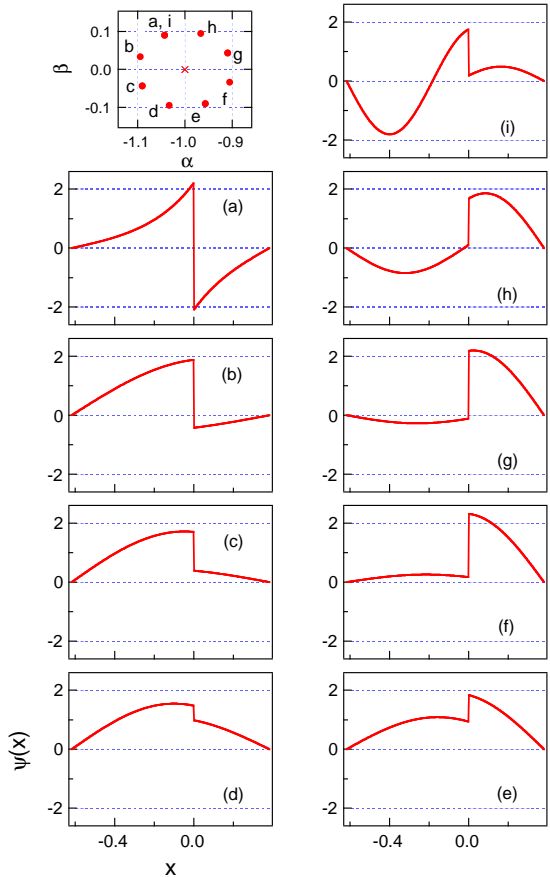

\label{fig4}
\center\psbox[hscale=0.45,vscale=0.45]{spfig4.epsf}
\caption{
The ground state is turned into the second
higher state when the parameters $(\alpha, \beta)$ trace
an adiabatic cycle around the singularity $(\alpha^* ,\beta^*)$
and come back to the original value.
}
\end{figure}
It is obvious that the structure of the parameter
space expressed by Eqs. (\ref{3}) and (\ref{2}) 
is the crucial element in bringing the double spiral structure
in energy surface and resulting wave-function aholonomy.
In mathematical term, the parameters 
$(\alpha , \beta , \gamma , \delta )$ span a $SL(2,{\bf R})$
manifold, which is isomorphic to the hyper-cylindrical manifold
$S^1 \times {\bf R}^2$.  The non-trivial structure of this manifold
as expressed in the non-zero homotopy group \cite{HU59} is 
a basis for the spiral aholonomy found in the current model.
The mathematical classification and generalization along this line
should open up a new vista
for the search of other systems with 
aholonomy.
It is worth pointing out, in this context, that certain gauge 
field theories possess the characteristics 
called {\it chiral anomaly} \cite{WI82} which can be understood 
in terms of the structure similar to Eq. (\ref{19}).

Finally, we would like to stress the potential utility of 
our findings in the fast-developing
nano-device technology.  In the coming age of quantum information 
processing and ``quantum mechanical engineering'', we
might even contemplate the actual applications of the
current scheme of wave-function morphology in manipulating the
wave functions to obtain the quantum states of desired 
shapes and energies. This might be achieved either 
in quantum dot settings as the potential problem of 
Eqs. (\ref{01}) - (\ref{1}), or in the heterojunction settings 
as discussed in Ref. \cite{BB95}.  

\vspace*{5mm}
We thank Prof. I. Tsutsui for enlightening discussions
on the topological aspects of our model.
We are grateful to Prof. T. Kawai and 
Prof. T. Shigehara for useful discussions.
Thanks are also due to Prof. D. Greene and 
Prof. L. Hunter for helpful suggestions.

\end{document}